\documentclass[a4paper,fleqn]{cas-dc}

\usepackage[numbers]{natbib}
\PassOptionsToPackage{table}{xcolor}
\usepackage{threeparttable}
\usepackage{tikz}

\def\tsc#1{\csdef{#1}{\textsc{\lowercase{#1}}\xspace}}
\tsc{WGM}
\tsc{QE}
\tsc{EP}
\tsc{PMS}
\tsc{BEC}
\tsc{DE}

\begin{document}
\let\WriteBookmarks\relax
\def\floatpagepagefraction{1}
\def\textpagefraction{.001}
\shorttitle{From Pixels to Pathology: Restoration Diffusion for Diagnostic-Consistent Virtual IHC}
\shortauthors{J. Liu et~al.}

\title [mode = title]{From Pixels to Pathology: Restoration Diffusion for Diagnostic-Consistent Virtual IHC}      



\author[1,3,6]{Jingsong Liu}[style=chinese,orcid=0009-0002-3174-3352]
\ead{jingsong.liu@tum.de}

\credit{Conceptualization, Methodology, Writing – original draft}


\author[4]{Xiaofeng Deng}[style=chinese]
\credit{Methodology, Validation}

\author[2,3]{Han Li}[style=chinese]
\credit{Validation, Writing – original draft}

\author[4,1,3]{Azar Kazemi}
\credit{Writing – review and editing, Visualization}
\author[1,3,6]{Christian Grashei}
\credit{Writing – review and editing, Formal analysis}
\author[1]{Gesa Wilkens}
\credit{Writing – review and editing}
\author[2,3]{Xin You}[style=chinese]
\credit{Writing – review and editing}
\author[1]{Tanja Groll}
\credit{Writing – review and editing}
\author[2,3]{Nassir Navab}
\credit{Writing – review and editing}
\author[1]{Carolin Mogler}
\credit{Writing – review and editing}
\author[1,3,6]{Peter J. Schüffler}
\credit{Conceptualization, Supervision, Funding acquisition, Writing – review and editing}
\ead{peter.schueffler@tum.de}
\cormark[1]

\cortext[cor1]{Corresponding author}


\affiliation[1]{organization={Institute of Pathology, Technical University of Munich, TUM School of Medicine and Health},
                city={Munich},
                country={Germany}}

\affiliation[2]{organization={Computer Aided Medical Procedures (CAMP), TU Munich},
                city={Munich},
                country={Germany}}

\affiliation[3]{organization={Munich Center for Machine Learning (MCML)},
                city={Munich}, 
                country={Germany}}
                
\affiliation[4]{organization={TUM School of Computation, Information and Technology},
                city={Munich}, 
                country={Germany}}
                
\affiliation[5]{organization={Department of Medical Informatics, School of Medicine, Mashhad University of Medical Sciences},
                city={Mashhad},
                country={Iran}}
                
\affiliation[6]{organization={Munich Data Science Institute (MDSI)},
                city={Munich}, 
                country={Germany}}

\begin{abstract}
Hematoxylin and eosin (H\&E) staining is the clinical standard for assessing tissue morphology, but it lacks molecular-level diagnostic information. In contrast, immunohistochemistry (IHC) provides crucial insights into biomarker expression, such as HER2 status for breast cancer grading, but remains costly and time-consuming, limiting its use in time-sensitive clinical workflows. To address this gap, virtual staining from H\&E to IHC has emerged as a promising alternative, yet faces two core challenges: (1) Lack of fair evaluation of synthetic images against misaligned IHC ground truths, and (2) preserving structural integrity and biological variability during translation. To this end, we present an end-to-end framework encompassing both generation and evaluation in this work. 
We introduce Star-Diff, a structure-aware staining restoration diffusion model that reformulates virtual staining as an image restoration task. By combining residual and noise-based generation pathways, Star-Diff maintains tissue structure while modeling realistic biomarker variability.
To evaluate the diagnostic consistency of the generated IHC patches, we propose the Semantic Fidelity Score (SFS), a clinical-grading-task-driven metric that quantifies class-wise semantic degradation based on biomarker classification accuracy. Unlike pixel-level metrics such as SSIM and PSNR, SFS remains robust under spatial misalignment and classifier uncertainty.
Experiments on the BCI dataset demonstrate that Star-Diff achieves state-of-the-art (SOTA) performance in both visual fidelity and diagnostic relevance. With rapid inference and strong clinical alignment, it presents a practical solution for applications such as intraoperative virtual IHC synthesis.
\end{abstract}

\begin{graphicalabstract}
\includegraphics{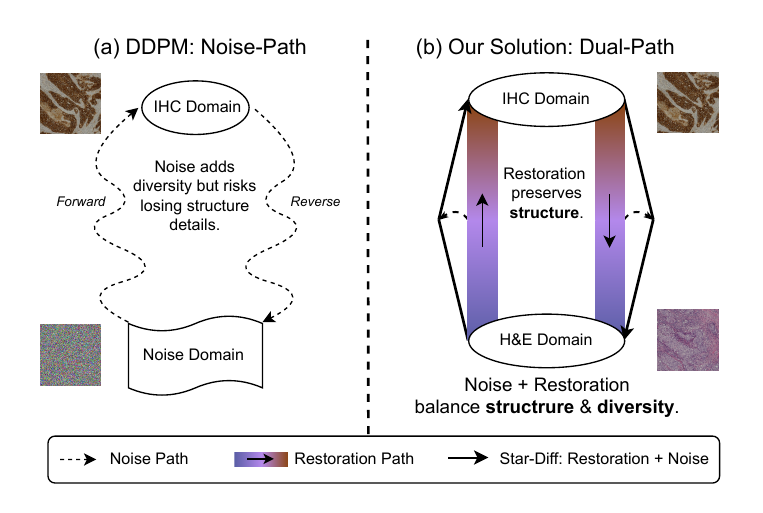}
\end{graphicalabstract}

\begin{highlights}
    \item Unlike prior translation-based models, we propose Star-Diff, a structure-diversity-balanced diffusion model that reformulates the task as image restoration. By introducing a deterministic restoration path alongside a stochastic noise path, Star-Diff achieves a controllable balance between H\&E structural details preservation and IHC molecular variability.
    
    \item To enable fair evaluation given the inevitable spatial misalignment between H\&E and IHC slides, we introduce the Semantic Fidelity Score (\textbf{SFS}), a classification-guided evaluation metric calibrated with class-wise performance degradation. Compared to traditional image quality metrics (e.g., SSIM, PNSR) that are highly sensitive to \textbf{spatial perturbations}, SFS delivers \textbf{stable evaluation scores} even under severe distortions such as translation, rotation, and deformation, making it particularly well-suited for histopathology staining tasks.

    \item We conducted thorough generalization experiments on the paired BCI dataset \cite{liu2022bci}, demonstrating Star-Diff's superior performance over 8 baselines in image quality metrics and also achieves the highest diagnostic relevance, exceeding the second-best model by over 5\% in diagnostic metrics. Additionally, we analyze the \textbf{interpretability} of Star-Diff using \textbf{saliency-based} visualizations, showing that it consistently focuses on diagnostically meaningful tissue regions during generation. Finally, we validate the \textbf{robustness} of the proposed SFS metric through spatial perturbation experiments, confirming its stability under misalignment and classifier bias, and establishing it as a clinically meaningful and robust assessment beyond pixel-level similarity.
\end{highlights}


\maketitle

\section{Introduction}
\label{sec:introduction}
Histopathological examination of Hematoxylin and Eosin (H\&E)-stained tissue slides is the clinical gold standard for diagnosing cancer. H\&E highlights cellular and morphological features
allowing pathologists to assess architectural patterns at cellular detail.
However, molecular biomarker information, such as expression levels of critical proteins, cannot be seen with the human eye in H\&E-stained slides. This can hinder diagnostic accuracy in cases that require biomarker-specific evidence \cite{magaki2018introduction}.
For further assessments, immunohistochemistry (IHC), an antibody-based staining method, was firstly proposed in the 1940s \cite{coons1942demonstration} to visualize the spatial expression levels of specific proteins, offering essential molecular cues for diagnosis, prognosis, and treatment selection.
Despite its utility, IHC is resource-intensive, both in terms of cost, processing time, and tissue, and may introduce tissue alignment inconsistencies due to sectioning and staining variability\cite{liu2025hasd, zhou2025utilizing}. As a result, in many low-resource settings or time-constrained workflows, pathologists are often restricted to H\&E slides, underscoring the need for computational approaches that can infer molecular information from standard H\&E staining \cite{anglade2020can}.

To mitigate the limitations of IHC staining  and enhance diagnostic accessibility, recent advances in deep learning have opened up new possibilities for inferring molecular information directly from H\&E slides. Several studies have demonstrated that certain biomarker expression patterns, although not explicitly visible to human observers in H\&E images, can be predicted with high accuracy using neural networks. For example, Farahmand et al. developed a convolutional neural network (CNN) to estimate HER2 scores in breast cancer based solely on H\&E images \cite{farahmand2022deep}, while Akbarnejad et al. leveraged vision transformers (ViTs) to predict ER, PR, and Ki-67 status, achieving area under the curve (AUC) scores approaching 0.90 across multiple biomarkers \cite{akbarnejad2023predicting}. These findings suggest that morphological features in H\&E are correlated with molecular profiles, implying a statistically learnable relationship between H\&E and IHC domains. This relationship can be modeled using deep generative frameworks that learn to synthesise corresponding IHC images conditioned on input H\&E images, which is commonly referred to as staining translation \cite{rivenson2019virtual}. By learning this mapping in a data-driven manner, generative models enable virtual biomarker visualization that is cost-effective, rapid, and particularly beneficial in resource-limited settings or high-throughput workflows where traditional IHC staining is impractical.
    While staining translation is promising, it presents key challenges. One major issue is the \textbf{structure-diversity tradeoff}: the generated IHC image should align with the source H\&E image to preserve spatial correspondence, yet emphasising structural fidelity can lead to under-representation of biomarker heterogeneity. Another challenge is \textbf{slide-pair spatial misalignment}: although paired H\&E and IHC slides originate from the same tissue block, they are cut at different depths, leading to misalignments that make pixel-wise comparisons unreliable. As a result, classical metrics like SSIM and PSNR often fail to capture the diagnostic relevance or biological fidelity of the virtual IHC (vIHC) images.

\begin{figure*}[t]
    \centering
    \centerline{\includegraphics[width=0.7\linewidth]{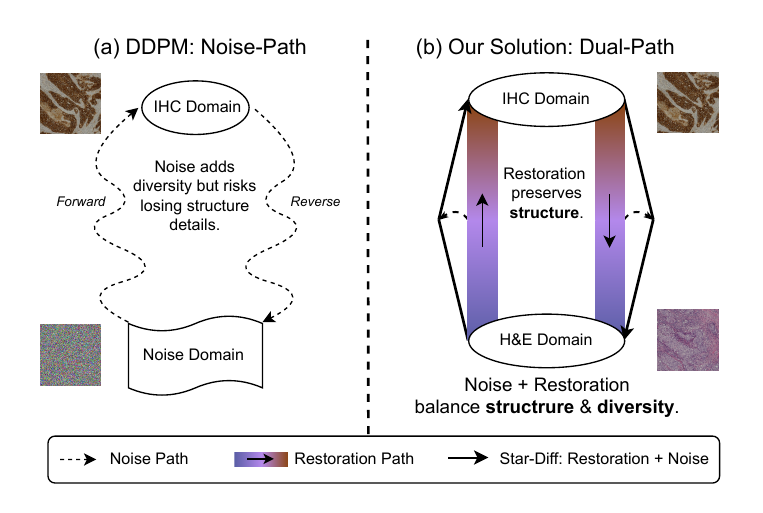}}
    \caption{Comparison of staining translation paradigms.
(a) The standard DDPM framework models staining translation from H\&E to IHC as an image translation problem and relies solely on the noise path, which introduces biological diversity but often results in high variance and structural inconsistency.
(b) Our approach reframes the problem as image restoration, treating the H\&E patch as the direct input. By introducing a restoration path from H\&E to IHC, our method effectively balances structural preservation and biological diversity.}
    \label{fig:overview}
\end{figure*}


To this end, we propose an integrated solution that combines a dual-path diffusion model, which leverages both restoration and noise pathways to balance tissue structure preservation with biomarker variability, and a task-driven evaluation metric designed to assess diagnostic consistency under inherent misalignment between H\&E and IHC images. Specifically, we propose a \textbf{sta}ining \textbf{r}estoration \textbf{diff}usion model \textbf{Star-Diff}. Unlike conventional approaches that treat staining translation as an image translation task, Star-Diff formulates it as an image restoration problem, which leverages a dedicated restoration path to deterministically preserve tissue architecture, as shown in Figure \ref{fig:overview}. The dedicated restoration path serves as continuous guidance from the source to the target domain. Along with the noisy path to introduce the randomness, the Star-Diff achieves the controllable balance between preserving tissue structure from H\&E slides and modelling biological variability in the vIHC images. In parallel, to enable fair and clinically relevant evaluation, we further propose the Semantic Fidelity Score (SFS), a classification-guided metric that remains robust to misalignment and classifier uncertainty. Specifically, we pretrained a ResNet-based classifier \cite{he2015deep} on real IHC images to predict biomarker expression from the generated outputs, providing a proxy for pathologist assessment. To further enhance the reliability, SFS is calibrated with class-wise performance degradation, offering a clinically meaningful and robust assessment beyond pixel-level similarity.
In summary, our contributions are three-fold:
\begin{itemize}
    \item Unlike prior translation-based models, we propose Star-Diff, a structure-diversity-balanced diffusion model that reformulates the task as image restoration. By introducing a deterministic restoration path alongside a stochastic noise path, Star-Diff achieves a controllable balance between H\&E structural details preservation and IHC molecular variability.
    %

    
    \item To enable fair evaluation given the inevitable spatial misalignment between H\&E and IHC slides, we introduce the Semantic Fidelity Score (\textbf{SFS}), a classification-guided evaluation metric calibrated with class-wise performance degradation. Compared to traditional image quality metrics (e.g., SSIM, PNSR) that are highly sensitive to \textbf{spatial perturbations}, SFS delivers \textbf{stable evaluation scores} even under severe distortions such as translation, rotation, and deformation, making it particularly well-suited for histopathology staining tasks.

    \item We conducted thorough generalization experiments on the paired BCI dataset \cite{liu2022bci}, demonstrating Star-Diff's superior performance over 8 baselines in image quality metrics and also achieves the highest diagnostic relevance, exceeding the second-best model by over 5\% in diagnostic metrics. Additionally, we analyze the \textbf{interpretability} of Star-Diff using \textbf{saliency-based} visualizations, showing that it consistently focuses on diagnostically meaningful tissue regions during generation. Finally, we validate the \textbf{robustness} of the proposed SFS metric through spatial perturbation experiments, confirming its stability under misalignment and classifier bias, and establishing it as a clinically meaningful and robust assessment beyond pixel-level similarity.

\end{itemize}

\section{Related Work}
\label{sec:related works}
~
We discuss existing approaches as categorized into the following three groups:

\subsection{Staining Translation via Color Mapping}
Early approaches of staining translation primarily focused on color normalization and mapping, treating it as a problem of statistical distribution alignment in color space. Reinhard et al. \cite{reinhard2001color} proposed a widely used technique that transfers the mean and standard deviation of image channels in the LAB space. This was later adapted for histology to reduce stain variability. Building on this, Macenko et al. \cite{macenko2009method} introduced a method using singular value decomposition (SVD) to estimate a stain matrix, while Vahadane et al. \cite{vahadane2016structure} improved stain separation using non-negative matrix factorization (NMF),  enabling more flexible and structure-preserving stain separation and transfer.
Although effective for visual consistency, these color-based methods fail to capture the complex, nonlinear relationships between staining types—particularly for antigen-specific stains like IHC. They may miss or distort critical pathological features, limiting their reliability for diagnostic use. This motivates the shift toward learning-based approaches that model deeper semantic relationships beyond color.

\subsection{Unpaired Staining Transfer}
To overcome the limitations of early color-based methods, semantic stain transfer methods have been developed. These approaches aim to ensure that the synthetic images not only match the target stain appearance but also preserve diagnostically critical features and tissue structures. A significant milestone in this direction is the introduction of CycleGAN \cite{zhu2017unpaired} enabling image-to-image translation using unpaired data. CycleGAN employs two generators and two discriminators to learn bidirectional mappings between source and target domains, with a cycle consistency loss to preserve the content of the input. This framework is particularly well-suited for histopathology, where paired HE–IHC samples are difficult to obtain. Building on CycleGAN, several works have proposed structure-aware and semantically guided adaptations for staining translation. For example, PC-StainGAN \cite{liu2021unpaired} takes advantage of a structural similarity constraint to preserve the structure during the translation. Other methods leverage auxiliary segmentation networks during training to enforce anatomical correctness in the generated stain images \cite{bouteldja2022improving}. ROIGAN \cite{boyd2022region} focuses translation efforts on diagnostically relevant regions, such as tumor or glomerular areas, guided by region-level supervision.

Most of these methods build on the CycleGAN framework due to its strong ability to learn mappings from unpaired data. However, they also inherit its limitations, such as training instability, mode collapse, and difficulty preserving fine structural details \cite{saad2024survey}. Besides, training with unpaired datasets further complicates the process, as the lack of pixel-wise alignment makes it challenging to ensure anatomical consistency. 

\subsection{Paired Staining Transfer}
These challenges have motivated a shift toward paired staining transfer approaches, where stronger supervision enables more accurate and structure-preserving translation. Recent research has focused on addressing imperfect spatial alignment between H\&E and IHC slides and enhancing the capture of clinically relevant features during translation.
A representative baseline is Pix2Pix \cite{isola2017image}, which employs a conditional GAN to learn a mapping from H\&E to IHC images. However, its reliance on strict pixel-wise supervision can be problematic due to inevitable misalignments. To mitigate this, multi-scale loss functions based on Gaussian pyramids have been introduced in \cite{liu2022bci} to promote consistency across different spatial resolutions, reducing sensitivity to fine-grained discrepancies.
To further enhance semantic guidance, BCI-Stainer \cite{zhu2023breast} incorporates biomarker classification as an auxiliary task. Features extracted from H\&E images are used to guide IHC synthesis, with a composite loss combining MAE, SSIM, and cosine similarity to balance structural fidelity and molecular relevance.
More recently, diffusion-based models such as PST-Diff \cite{he2024pst} have demonstrated superior performance in generating high-quality and diverse IHC images. By introducing both structural and pathological consistency constraints, these models better preserve diagnostic information and offer improved training stability compared to traditional GAN-based methods.
Despite progress in virtual staining, existing methods often fall short in two key aspects: they do not explicitly address the dual challenge of structure-preserving and variability-aware staining translation from H\&E to IHC, and they rely on evaluation strategies that fail to reflect clinical utility, often emphasizing pixel-level similarity over diagnostic relevance.\\

Unlike prior translation-based models, we propose \textbf{Star-Diff} model, which introduces a deterministic restoration path alongside a stochastic noise path, to preserve H\&E structural details while modeling IHC molecular variability.

\section{Methods}
\label{sec:methods}
~
\begin{figure*}[t]
    \centering
    \includegraphics[width=0.8\linewidth]{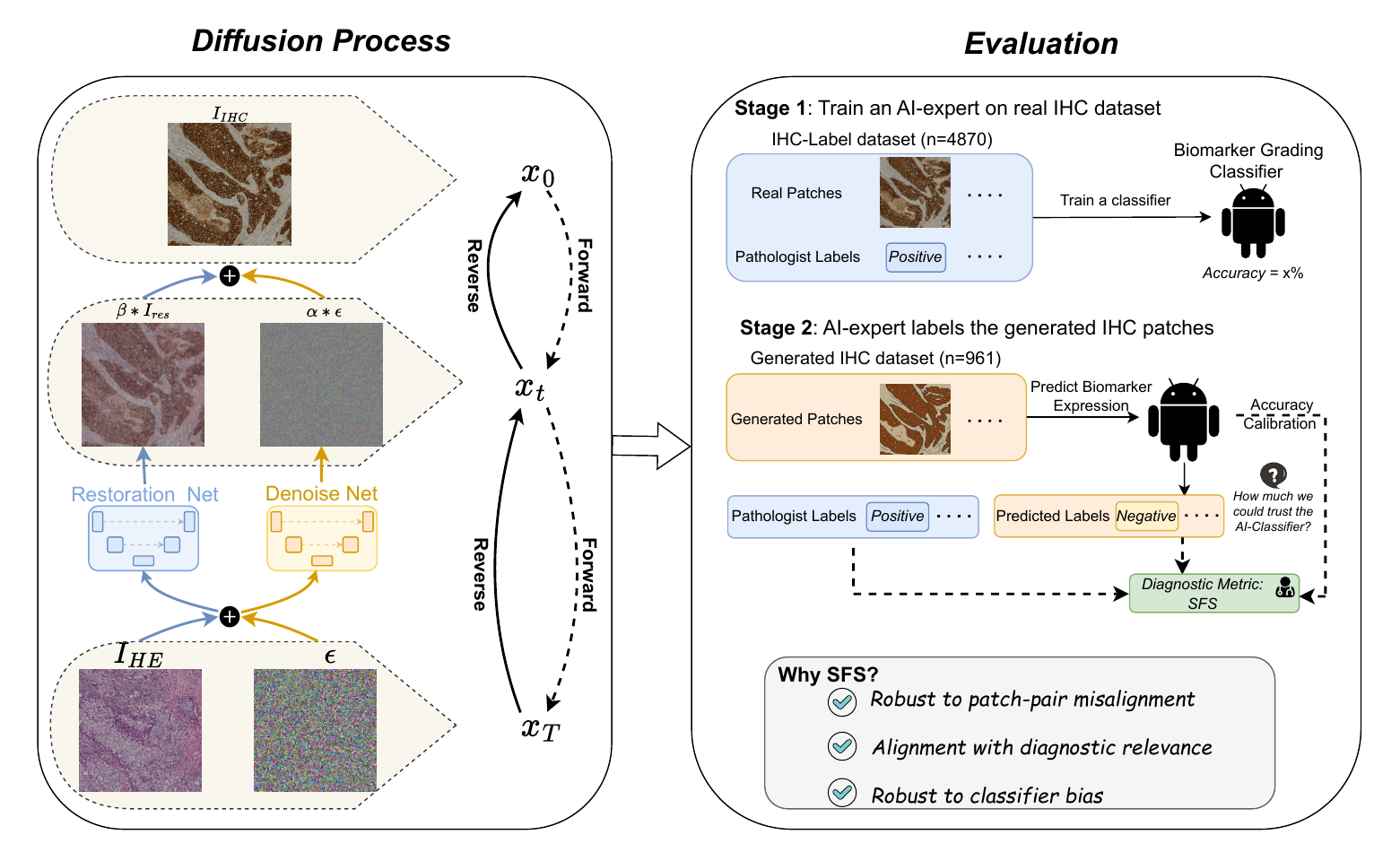} 
    \caption{We propose Star-Diff, a structure-diversity-balanced diffusion framework that formulates staining translation as a restoration task. (Left) The denoising process integrates restoration and noise prediction to reconstruct semantically faithful IHC images from perturbed H\&E input. (Right) To provide a clinically meaningful and misalignment-robust assessment of generated IHC images, we introduce a two-stage classifier-based evaluation strategy: (1) an AI expert is trained on real IHC data with pathologist annotations; (2) the trained expert predicts the labels of vIHC images. The Semantic Fidelity Score (SFS) measures alignment between AI and pathologist labels while calibrating for classifier reliability. Unlike traditional metrics such as SSIM or PSNR, SFS reflects semantic preservation and is robust to patch misalignment and classifier bias.}
    \label{fig:fig2}
\end{figure*}

In this section, we first define the staining translation mathematically and revisit the analytical solution with the diffusion model. Then, we describe our key contribution of the Star-Diff model, which approaches the staining translation as an image restoration problem by adding a dedicated restoration path to keep the balance between structure preservation and staining variability. Finally, we discuss the weakness of existing evaluation metrics for staining translation tasks and introduce our clinical-tasks-based evaluation metric SFS. The novel integrated framework is illustrated in Figure \ref{fig:fig2}.

\subsection{Problem definition}

Given a set of H\&E-stained images $I_{\text{he}}$ and their corresponding IHC-stained images $I_{\text{ihc}}$, the goal of staining translation is to learn a mapping function $f: \mathcal{X}_{\text{he}} \rightarrow \mathcal{X}_{\text{ihc}}$ that generates IHC images which are both structurally consistent with the input H\&E image and biologically meaningful in terms of biomarker expression. Here, $\mathcal{X}_{\text{he}}$ and $\mathcal{X}_{\text{ihc}}$ denote the underlying spaces of H\&E and IHC images, respectively, and $I_{\text{he}} \subset \mathcal{X}_{\text{he}}$, $I_{\text{ihc}} \subset \mathcal{X}_{\text{ihc}}$ represent the datasets used for training.

In the context of conditional diffusion models \cite{ho2020denoising}, this mapping can be interpreted as generating $I_{\text{ihc}}$ by reversing a noise-adding process. Let $x_0 \in  I_{\text{ihc}}$ be the target image and $x_T \sim \mathcal{N}(0, I)$ be Gaussian noise. A forward diffusion process progressively corrupts $x_0$ into $x_T$ through a Markov chain from $t = 1, \ldots, T$:
\begin{equation}
q(x_t \mid x_{t-1}) = \mathcal{N}(x_t; \sqrt{\alpha_t} x_{t-1}, (1-
\alpha_t) I)
\label{eq:forward_process}
\end{equation}

where $\alpha_t$ is a predefined noise schedule. The equation (\ref{eq:forward_process}) could also be written as 
\begin{equation}
x_t = \sqrt{\bar{\alpha}_t}x_0 + \sqrt{1-\bar{\alpha}_t}\epsilon 
\label{eq:resampling}
\end{equation}
where $\bar{\alpha}_t = \prod_{s=1}^{t}(\alpha_s)$ is the cumulative product of the noise schedule coefficients, and $\epsilon \sim \mathcal{N}(0, I)$ is standard Gaussian noise. 
The generative task then becomes learning the reverse process $p_\theta(x_{t-1} \mid x_t, I_{\text{he}})$ conditioned on the input H\&E image.

Our goal is to model this conditional generation process such that the output $\hat{x}_0$ not only visually resembles the real IHC image but also retains the structural layout of $I_{\text{he}}$ and reflects plausible biomarker variability.



\subsection{IHC Generation with Restoration Guidance}
While standard diffusion models like DDPM are effective for generative tasks, they lack explicit structural constraints during the reverse process, which can lead to the loss of critical tissue architecture in histopathological images \cite{ho2020denoising}. To address this limitation, and inspired by recent advances \cite{liu2024residual, xia2023diffir, li2022srdiff}, we propose to reformulate staining translation as a \textbf{structure-aware IHC restoration task}, where the residual between the H\&E and IHC domains serves as a deterministic guidance signal throughout the diffusion process.

\textbf{Forward process.} Specifically, we define the restoration image as:
\begin{equation}
I_{\text{res}} = I_{\text{ihc}} - I_{\text{he}}
\label{eq:restoration}
\end{equation}

In contrast to standard DDPMs, where only random Gaussian noise is added during the forward process, we incorporate an additional deterministic restoration schedule. Similar to Equation~\ref{eq:resampling}, the noisy sample at timestep $t$ is defined as:
\begin{equation}
x_t = x_0 + \bar{\alpha}_t \epsilon + \bar{\beta}_t I_{\text{res}}
\label{eq:sampling_res}
\end{equation}
where $\epsilon \sim \mathcal{N}(0, I)$, $\bar{\alpha}_t$ controls the stochastic noise level, and $\bar{\beta}_t$ is a predefined restoration schedule that deterministically integrates structural guidance from the restoration.

\textbf{Reverse process.} In our framework, the forward process perturbs the target IHC image $x_0$ by gradually adding Gaussian noise and a deterministic restoration signal. To sample the target IHC image, we train two networks in parallel:
\begin{itemize}
    \item A \textbf{restoration prediction network} $r_\theta(x_t, t, I_{\text{he}})$, which estimates the residual component $\hat{I}_{\text{res}}$ between the H\&E and IHC images. This network guides the reverse process by enforcing structural consistency with the input H\&E image.
    
    \item A \textbf{noise prediction network} $\epsilon_\theta(x_t, t, I_{\text{he}})$, which predicts the noise added during the forward process. This network is responsible for modelling biomarker variability and accounts for the stochasticity in the restoration.
\end{itemize}
Given the outputs of the two networks, the reverse sampling distribution is defined as:
\begin{equation}
p_\theta(x_{t-1} \mid x_t) := q_\sigma(x_{t-1} \mid x_t, \epsilon_\theta, r_\theta)
\label{eq:reverse_prob}
\end{equation}
where $q_\sigma$ is the transfer probability combining both restoration guidance and noise estimation.
The actual sampling at timestep $t$ is then computed as:
\begin{equation}
x_{t-1} = x_t - \gamma_t {r}_{\theta} - \eta_t \epsilon_\theta
\label{eq:reverse_sampling_simplified}
\end{equation}
where $\gamma_t$ and $\eta_t$ control the balance between deterministic structural guidance from the H\&E–IHC restoration and the variability introduced by the learned noise correction. This formulation decouples anatomical structure preservation from stochastic uncertainty, enabling the model to generate biomarker-aware IHC images that are both spatially consistent and clinically plausible.

\textbf{Connection to DDPM.} Our framework reduces to the standard DDPM formulation when the restoration guidance is disabled, i.e., $\gamma_t = 0$. 

%

\subsection{Novel clinical-tasks-based evaluation strategy}

\textbf{Weakness of existing metrics} Existing evaluation metrics for staining translation, such as the structural similarity index measure (SSIM) \cite{wang2004image}, the peak signal-to-noise ratio (PSNR)\cite{hore2010image} and the mean square error (MSE), focus on pixel-level similarity between generated and ground-truth images. While effective for natural images with perfect alignment, they become unreliable in histopathology due to frequent spatial misalignments between H\&E and IHC slides.
In practice, adjacent tissue sections often show deformation, rotation, or cutting artifacts \cite{liu2022bci, lotz2023comparison}, making precise pixel-to-pixel comparison unrealistic. As shown in Fig.~\ref{fig:pertubation}, even small shifts can drastically lower SSIM and PSNR scores, despite the preservation of diagnostic characteristics.
Relying solely on these metrics can unfairly penalize biologically meaningful results. This underscores the need for evaluation strategies that assess clinical relevance rather than strict pixel-level agreement.

\textbf{Novel misalignment-robust evaluation strategy} The primary objective of staining translation is to assist clinical decision-making by generating diagnostically meaningful IHC images. Instead of relying solely on pixel-level metrics, we propose a task-driven evaluation strategy grounded in a clinically relevant downstream task, focusing on global semantic information rather than local pixel-level comparisons, and offering greater robustness to spatial misalignment.

To this end, we train a ResNet-based classifier \cite{he2015deep} to predict biomarker expression from real IHC images, using pathologist-verified annotations.
Once trained, this classifier is applied to the virtual IHC (vIHC) images, and its performance serves as a proxy for evaluating the semantic consistency between vIHC and IHC domains. 
However, since the classifier only approximates pathologist-level interpretation, biases could be introduced during training, such as underfitting and overfitting, which may affect its reliability \cite{kelly2019key}. These factors can degrade its performance on IHC images and introduce noise into the evaluation of vIHC.

To address this limitation, SFS is calibrated with raw accuracy bias by explicitly accounting for class-wise recall degradation. This allows for a more semantically grounded comparison of translation quality between methods, even when classifier performance is sub-optimal.

Let $C$ denote the number of biomarker classes. For each class $c \in \{1, \dots, C\}$, we compute the recall on real and generated images respectively as:
\begin{equation}
R_c^{\mathrm{real}} = \frac{TP_c^{\mathrm{real}}}{N_c}
\label{eq:recall_real}
\end{equation}

\begin{equation}
R_c^{\mathrm{gen}} = \frac{TP_c^{\mathrm{gen}}}{N_c}
\label{eq:recall_gen}
\end{equation}
where \( TP_c^{\mathrm{real}} \) and \( TP_c^{\mathrm{gen}} \) denote the number of real and generated images correctly classified as class \( c \), and \( N_c \) is the total number of images with ground truth label \( c \).



We define the average semantic degradation across all classes as:
\begin{equation}
\mathrm{AvgDeg} = \frac{1}{C} \sum_{c=1}^{C} \left( R_c^{\mathrm{real}} - R_c^{\mathrm{gen}} \right)
\label{eq:avg_deg}
\end{equation}
Let $\mathrm{Acc}_{\mathrm{gen}}$ denote the overall classification accuracy on generated images. The \textit{Semantic Fidelity Score (SFS)} is defined as:

\begin{equation}
\mathrm{SFS} = \frac{ \mathrm{Acc}_{\mathrm{gen}} + \left(1 - \mathrm{AvgDeg} \right) }{2}
\label{eq:sfs}
\end{equation}

This metric ranges from 0 to 1, where higher scores indicate that the vIHC images preserve diagnostic information aligned with real data, even under imperfect spatial alignment or mild classifier uncertainty.

\section{Experiments and Results}
\label{sec:experiments}
~
\definecolor{diffusionbg}{RGB}{230, 245, 255}
\newcommand{\bluebox}{\tikz\draw[blue,fill=diffusionbg] (0,0) rectangle (0.15,0.15);}
\newcommand{\reddown}{\textcolor{red}{$\downarrow$}}

We rigorously evaluate Star-Diff and compare it to alternative approaches. Besides the quality metrics SSIM and PNSR that are impacted by the poor alignment quality, we also assess the diagnostic consistency of the vIHC images using diagnostic-guided metrics accuracy and SFS. To explore the explainability of the Star-Diff generation process, we further visualize attention maps across diffusion reverse paths. Finally, we perform perturbation experiments to assess the robustness of the proposed SFS metric against spatial misalignment and classifier bias. All experiments were conducted on an NVIDIA GeForce RTX 3090 GPU.

\subsection{Staining Translation experimental design}
\textbf{Dataset.} 
For evaluation, we use the publicly available BCI challenge dataset \cite{liu2022bci} containing 4,870 paired H\&E and HER2-stained image patches from 51 whole-slide image pairs of breast cancer cases. HER2 is a clinically relevant biomarker for breast cancer diagnosis, with expression levels manually annotated as 0, 1+, 2+, or 3+. Annotations are provided at the slide level, meaning patch-level labels may exhibit intra-slide variability. Furthermore, although H\&E and IHC slides originate from the same tissue block, they are co-registered at slide level rather than pixel-level, leading to potential spatial misalignments between presumably corresponding patches. We follow the train/test split provided in the BCI challenge \cite{liu2022bci}. Their training split is further divided into 80\% for training and 20\% for validation.

\textbf{Metrics.} 
As suggested in the BCI challenge, \textbf{SSIM} and \textbf{PSNR} are used to evaluate the pixel-level similarity between generated and reference IHC images. In addition, we follow the challenge's protocol by defining an \textbf{overall quality ranking} as: $0.6 \times \text{SSIM Rank} + 0.4 \times \text{PSNR rank}$ \cite{liu2022bci}.

To assess diagnostic consistency, we further evaluate the vIHC using \textbf{accuracy} and our proposed \textbf{SFS}, which captures class-wise semantic alignment with ground-truth biomarker expression.
To compute SFS, we train a ResNet-based classifier on the training split of IHC patches from the BCI Challenge to predict biomarker expression, achieving over 86\% accuracy on the test split. For evaluation, HER2 scores are binarized into two clinically meaningful categories: \textit{HER2-positive} (2+ and 3+) and \textit{HER2-negative} (0 and 1+). The classifier is then applied to the whole virtual IHC (vIHC) image set to obtain prediction accuracy. Meanwhile, SFS is further computed by combining the overall classification accuracy with class-wise recall degradation. This provides a more clinically relevant measure of the translation quality, beyond simple pixel-level similarity.

\textbf{SOTA methods.} 
We compare our method with representative staining translation approaches, including traditional color normalization, unsupervised learning, and supervised generative models. \textbf{Color normalization methods} include Reinhard Normalization \cite{reinhard2001color} and Macenko Normalization \cite{macenko2009method}, which align color distributions between source and target domains using handcrafted transformations. \textbf{Unsupervised approaches} are represented by CycleGAN \cite{zhu2017unpaired}, which learns bidirectional mappings between H\&E and IHC domains without requiring paired data. \textbf{Supervised models} include Pix2Pix \cite{isola2017image}, which employs conditional GANs with L1 loss; Pix2Pix-Pyramid \cite{liu2022bci}, which extends Pix2Pix with multi-scale Gaussian pyramid losses to improve structural consistency; PST-Diff \cite{he2024pst}, a diffusion-based method incorporating structural and pathological constraints; and Palette \cite{saharia2022palette}, a comprehensive DDPM based framework for image translation.

 For baseline models, CycleGAN and Pix2Pix are implemented and fine-tuned on the BCI training set, and we report their best-performing checkpoints based on validation performance.
For PST-Diff, we adopted the result from the original paper directly, since the weights are not released.
All models are evaluated on the held-out BCI test set using the metrics described above, including SSIM, PSNR, accuracy and SFS.

\subsection{Staining Translation results}
Table~\ref{tab:results_bci} summarizes the quantitative performance of various staining translation methods on the BCI test set. We evaluate image quality using PSNR and SSIM, and assess clinical relevance using classification accuracy and SFS. For stochastic diffusion-based models, we perform three independent sampling runs and report the mean and standard deviation.

\textbf{High-Quality Image Generation.} %
Star-Diff achieves state-of-the-art image quality, outperforming both GAN-based and diffusion-based baselines in PSNR and SSIM. Unlike classical color mapping or unpaired translation methods, which struggle with structural fidelity, Star-Diff redefines staining translation as a restoration problem rather than conventional translation, and introduces restoration guidance to preserve tissue structure explicitly, together enabling visually accurate and structurally consistent IHC image generation.

\textbf{Enhanced Diagnostic Fidelity.} %
Diffusion models, while slightly lagging behind GANs in pixel-level metrics, generally achieve stronger performance in diagnostic evaluations, reflecting their ability to model plausible distributions of biomarker expression. Among them, Star-Diff stands out—its restoration guidance enhances distribution modeling and leads to superior performance in both Accuracy and SFS metrics.


\textbf{Reducing Uncertainty for Improved Clinical Reliability.}
Star-Diff exhibits lower variance in PSNR and SSIM compared to other diffusion baselines, demonstrating its ability to balance biological diversity with structural consistency. This stability stems from its restoration-guided design, which anchors the denoising process to the input H\&E structure while introducing controlled variability. 

\textbf{Interpretation of SOTA Methods Performance}
Different SOTA methods exhibit distinct performance patterns, as shown in Table~\ref{tab:results_bci}.
The color mapping baselines (Reinhard, Macenko, and Vahadane) statistically align color distributions from H\&E to IHC patches but fail to capture structural details or biomarker expression patterns. As a result, they achieve poor performance across both image quality and diagnostic metrics.
Unsupervised methods like CycleGAN and CUT, where unpaired translation models are trained adversarially with regularization, offer marginal improvements over color mapping but still lack the capacity to preserve structural fidelity or biomarker information effectively.
Supervised paired methods significantly outperform unsupervised approaches. GAN-based models such as Pix2Pix and Pix2Pix-Pyramid achieve strong PSNR and SSIM scores due to direct pixel-level supervision during training. However, they tend to focus more on local structural properties while overlooking global diagnostic information.
Diffusion models such as Palette and PST-Diff, while slightly underperforming supervised GANs in pixel-based metrics, demonstrate stronger results in diagnostic metrics. For example, Palette achieves higher accuracy and SFS than Pix2Pix-Pyramid (Acc: 0.621 vs. 0.610, SFS: 0.688 vs. 0.687), highlighting the diffusion models’ ability to model the bidirectional mapping between source and target distributions. Additionally, the diversity of the generated IHC patches reflects the natural variability in staining.
Nonetheless, due to the lack of direct supervision between generated and target images, these DDPM-based models suffer from greater structural inconsistency and higher variance across samples. PST-Diff, in particular, demonstrates substantial variability in PSNR and SSIM, stemming from the inherent randomness of the denoising process.

Star-Diff reinterprets the staining translation task as an \textbf{image restoration problem} by introducing a direct residual path. This residual guidance, together with the noise path, allows for a controlled balance between structural preservation and staining variability. As a result, Star-Diff establishes new SOTA performance across both image quality and diagnostic relevance metrics.


\begin{table*}[t]
\centering
\caption{Comparison of staining translation methods on the BCI dataset. Best results in each column are highlighted in bold, and the second-best results in each column are underlined.}
\label{tab:results_bci}
\begin{threeparttable}
\begin{tabular}{l|cc|c|cc}
\hline
\multirow{2}{*}{\textbf{Method}} & \multicolumn{3}{c|}{\textbf{Image Quality Metrics}} & \multicolumn{2}{c}{\textbf{Diagnostic Metrics}} \\
\cline{2-6}
 & PSNR (dB)$\uparrow$ & SSIM$\uparrow$ & Quality Rank$\downarrow$ & Accuracy$\uparrow$ & SFS$\uparrow$ \\
\hline

\textbf{Color Mapping} &&&&\\

Reinhard\cite{reinhard2001color}      & 15.34 & 0.44 &8th& 0.60 & 0.65 \\
Macenko\cite{macenko2009method}        & 15.49 & 0.41 &6th& 0.57 & 0.63\\
Vahadane\cite{vahadane2016structure}     & 15.04 & 0.35 &9th & 0.59 & 0.67\\

\hline
\textbf{Unpaired Supervised} &&&&\\
CycleGAN\cite{zhu2017unpaired}              & 16.20 & 0.37 & 7th & 0.59 & 0.65 \\
\hline
\textbf{Paired Supervised} &&&&\\
Pix2Pix*\cite{isola2017image}            & 19.63 & 0.42 &4th & 0.60 & 0.67 \\
Pix2Pix-Pyramid*\cite{liu2022bci}       & \textbf{21.61} & 0.48 & \underline{2nd} & 0.61 & 0.69 \\
\rowcolor{diffusionbg}
Palette\cite{saharia2022palette}  & $17.13\pm0.53$ &\underline{$0.53\pm0.08$} &3rd & \underline{$0.62\pm0.05$} &\underline{$0.69\pm0.03$} \\
\rowcolor{diffusionbg}
PST-Diff$^{\dagger}$\cite{he2024pst}  &$16.75\pm4.20$ & $0.38\pm0.11$ &5th
&- & - \\
\hline
\rowcolor{diffusionbg}
\textbf{Star-Diff (Ours)}                  & \underline{$21.30\pm0.01$} & \textbf{\boldmath$0.53 \pm 0.00$} & \textbf{1st} &
\textbf{\boldmath$0.68 \pm 0.02$}& \textbf{\boldmath$0.74 \pm 0.01$}\\

\end{tabular}
\begin{tablenotes}
\item[*]  Results obtained from \cite{liu2022bci}.
\item[$\dagger$] Results obtained from \cite{he2024pst}.
\item[\bluebox] Highlights diffusion models.
\end{tablenotes}
\end{threeparttable}
\end{table*}


\subsection{Explainability of generation process}
\begin{figure}[t]
    \centering

    \includegraphics[width=\linewidth]{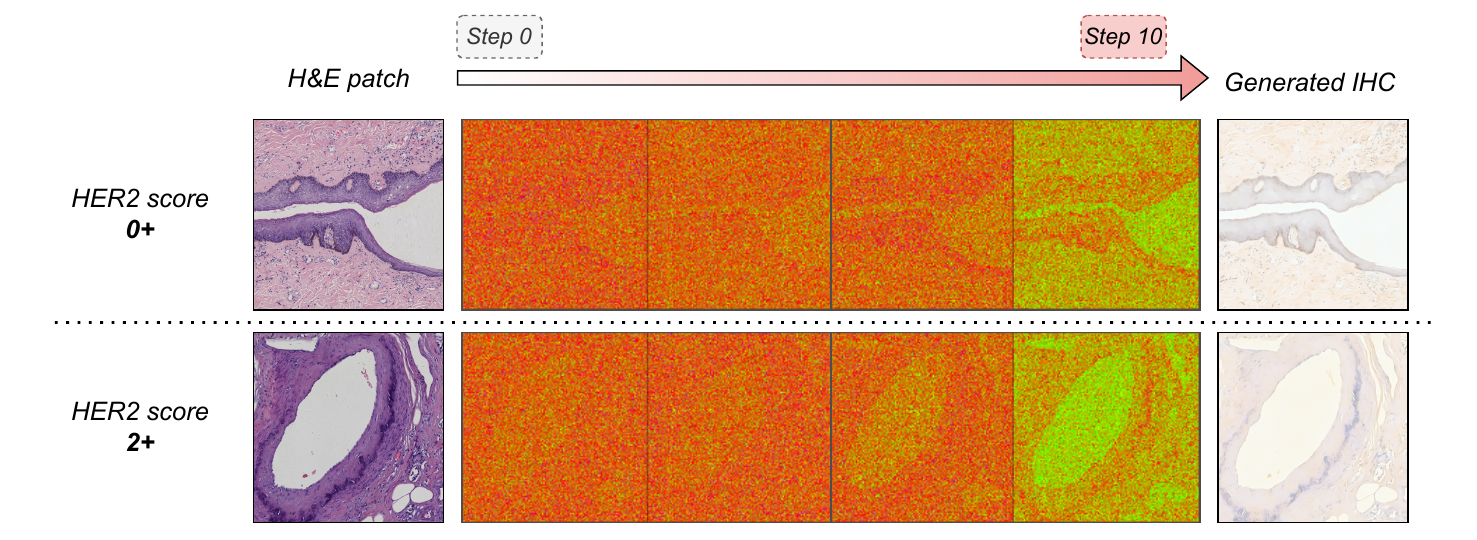} 
    \caption{Saliency visualization using RISE during the denoising process. We selected the most representative breast cancer patches with HER2 scores of 2+ and 0+ for visualization. As denoising progresses, the model’s attention shifts toward the stained regions, aligning with diagnostically meaningful tissue structures.}
    \label{fig:rise_saliency}
\end{figure}

We employ explainable AI (xAI) to understand how our model maintains structural and semantic fidelity. We adapt RISE \cite{petsiuk2018rise}, a black-box saliency method, to visualize model attention throughout the denoising process. RISE estimates pixel importance by probing the model with randomly masked inputs and measuring their influence on the output. In the resulting saliency maps (Fig.~\ref{fig:rise_saliency}), red pixels indicate high attribution (greater influence), while green pixels indicate low attribution.
The model progressively focuses on stained tissue regions in the H\&E input, especially during the later denoising steps when structural details become more visible. Red regions tend to align with areas of higher biomarker expression, whereas non-tissue or background regions show low attribution, suggesting minimal influence on generation.

This behavior supports our objective of producing diagnostically meaningful outputs. Notably, early denoising steps exhibit more diffuse and uncertain attention due to high noise levels, but progressively shift attention toward critical biomarker expression structures as noise is reduced. These observations confirm that the restoration-guided denoising process encourages anatomically and diagnostically informed generation, avoiding shortcuts like overemphasizing background regions with high structural similarity but low clinical relevance.

\subsection{Robustness of SFS}

\textbf{Robustness to Spatial Misalignment.}
To evaluate the robustness of different evaluation metrics under spatial misalignment between source and target image pairs, we conducted a perturbation analysis using \textbf{identical IHC patches} as the baseline. Three common spatial perturbations were applied: translation, rotation, and elastic deformation. The performance drop of each metric was then measured after each perturbation.

As shown in Table~\ref{tab:perturbation_analysis}, the traditional image quality metrics SSIM and PSNR are \textbf{highly sensitive} to minor spatial variations. For example, a 5px translation leads to a 47.9\% drop in SSIM, despite the diagnostic content remaining unchanged. Similarly, small rotations of 5–15° result in over 51\% decrease in SSIM. These findings underscore the limitations of classical image quality metrics that heavily rely on pixel-level alignment and may not accurately reflect clinically relevant features.

In contrast, the diagnostic metrics Accuracy and SFS remain considerably stable across all perturbation types. Notably, even under severe elastic deformation, the drops in Accuracy and SFS are limited to 3.4\% and 2.1\%, respectively. This demonstrates their robustness to the common spatial misalignment between H\&E and IHC patches and suggests that they are better suited for evaluating staining translation in terms of preserving diagnostic relevance.

\textbf{Robustness to Classifier Bias.}
To evaluate the robustness of SFS to classifier bias, we simulate three levels of classifier reliability: \textit{underfit}, \textit{properly-fit}, and \textit{overfit}. We train the model for a total of 60 epochs and monitor performance on both train and test splits to identify different stages of model fitting:

\begin{itemize}
\item \textit{Underfit (Epoch 20):} The classifier exhibits low accuracy on both training and test sets, indicating it has not yet learned meaningful patterns.
\item \textit{Properly-fit (Epoch 40):} The classifier achieves high accuracy on both sets, demonstrating good generalization.
\item \textit{Overfit (Epoch 60):} While the classifier reaches near-perfect accuracy on the training set, its performance on the test set deteriorates, signaling overfitting.
\end{itemize}

\begin{figure}[t]
    \centering
    \includegraphics[width=\linewidth]{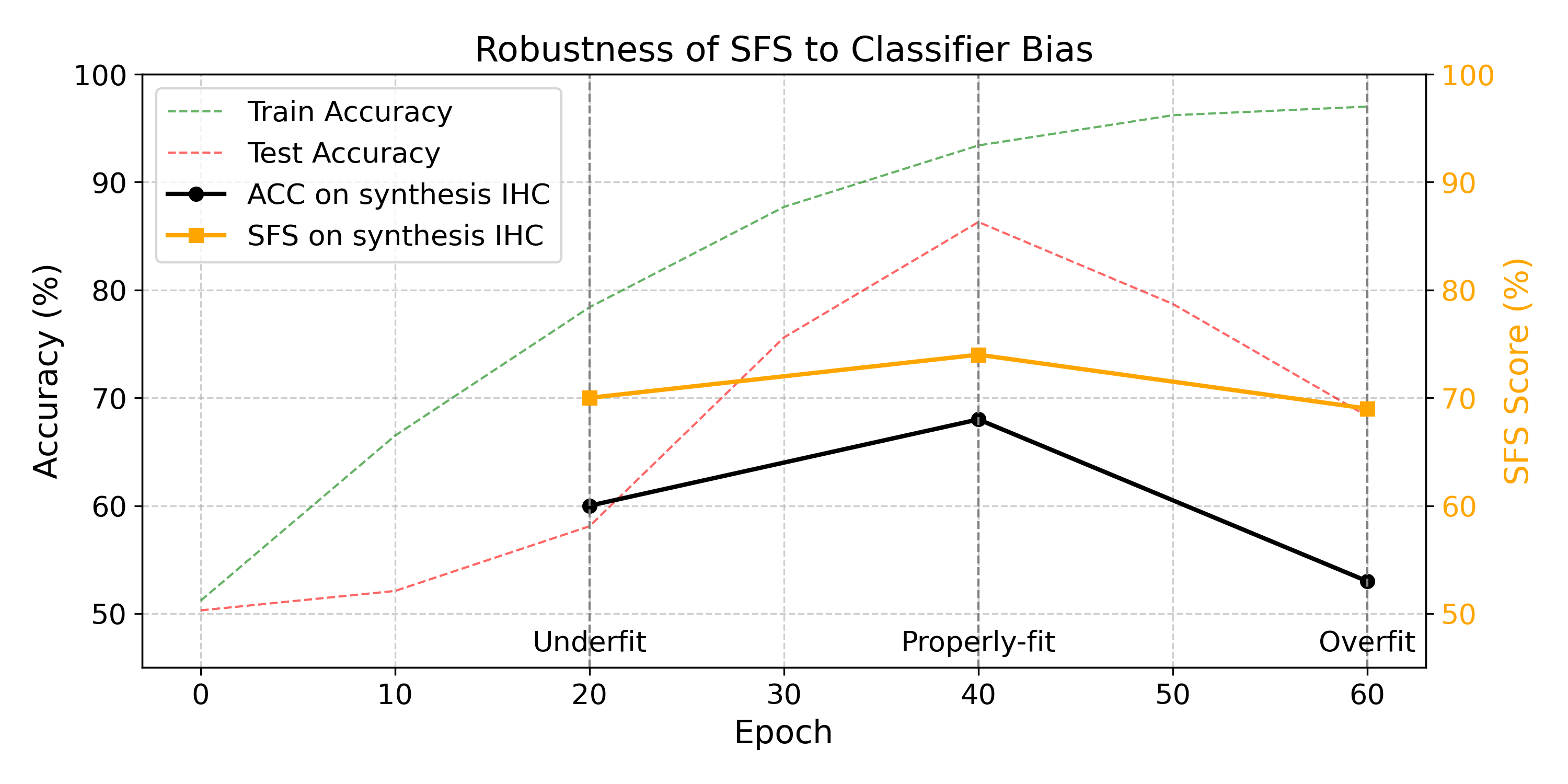}
    \caption{The training and test accuracy curves show how classifier performance evolves over 60 epochs, highlighting three key stages: underfitting (low accuracy on both sets), proper fitting (high accuracy on both), and overfitting (high training but declining test accuracy). Accuracy and SFS are measured on synthetic IHC data at each stage. While accuracy drops significantly during underfitting and overfitting, SFS remains comparatively stable, demonstrating greater robustness to classifier bias.}
    \label{fig:sfs_robustness}
\end{figure}

\begin{figure}[t]
    \centering
    \includegraphics[width=\linewidth]{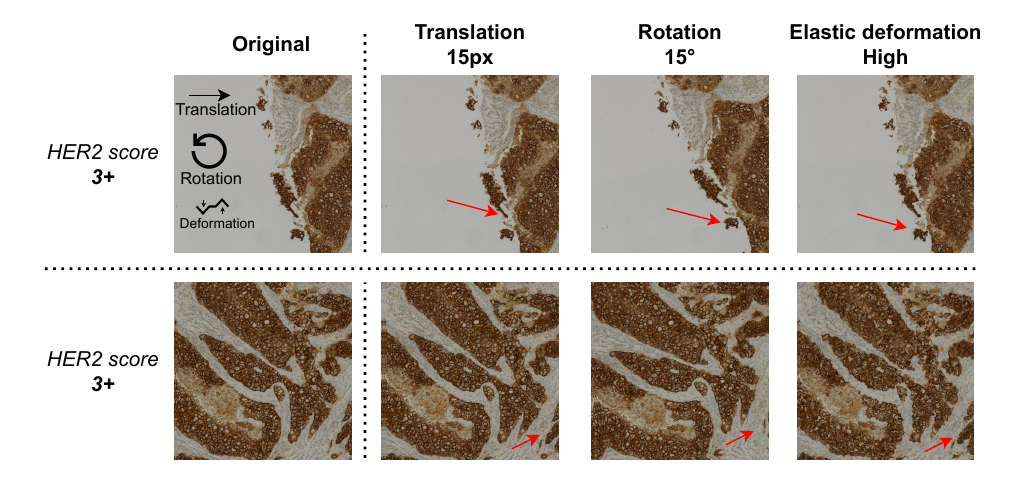} 
    \caption{Examples of visual perturbations applied to IHC patches with breast cancer HER2 score 3+. The first column shows the original patches, while the following columns demonstrate the effects of three perturbations: translation (15px), rotation (15°), and high elastic deformation. These spatial distortions visibly alter structural alignment but cause minimal change to the underlying semantic content. }
    \label{fig:pertubation}
\end{figure}

For each of these stages, we freeze the classifier and evaluate \textbf{the vIHC images} using both Accuracy and SFS. As shown in Figure~\ref{fig:sfs_robustness}, Accuracy is highly sensitive to classifier quality, dropping sharply in the overfitting scenario due to poor generalization. In contrast, SFS remains comparatively stable across all settings, as it is calibrated to account for variations in classifier performance. This demonstrates that SFS is a reliable indicator of semantic consistency in generated IHC images, even when the evaluation classifier is imperfect or biased. 

\begin{table*}[htbp]
\centering
\caption{Perturbation Analysis Results. Performance drops are computed relative to the unperturbed baseline.}
\label{tab:perturbation_analysis}
\begin{threeparttable}
\begin{tabular}{l|cc|cc|ccc}
\hline
\multirow{2}{*}{\textbf{Perturbation}} & \multicolumn{2}{c|}{\textbf{Image Quality Metrics $\uparrow$}} & \multicolumn{2}{c|}{\textbf{Diagnostic Metrics $\uparrow$}} & \multicolumn{3}{c}{\textbf{Performance Drop $\downarrow$ (\%)}}  \\
\cline{2-8}
&SSIM & PSNR (dB) & Accuracy  & SFS &  SSIM &Accuracy & SFS \\
\hline
\textbf{Unpertubated baseline} & & & & \\
Identical IHC pair & 1.00 & Inf & 0.87 &0.94 & - & - & -\\
\hline
\textbf{Translation Perturbations} & & & & & & & \\
5px & 0.52 (0.48\reddown) & 25.13 &\textbf{ 0.86} (\textbf{0.01}\reddown) & \textbf{0.93} (\textbf{0.01}\reddown) &47.9& 1.1  & \textbf{1.0} \\
10px & 0.49 (0.51\reddown) & 24.16 & \textbf{0.86} (\textbf{0.01}\reddown) & \textbf{0.93} (\textbf{0.01}\reddown) &51.0 & 1.1 & \textbf{1.0 }\\
15px & 0.49 (0.51\reddown) & 23.66 & \textbf{0.86} (\textbf{0.01}\reddown) &\textbf{0.93} (\textbf{0.01}\reddown)& 51.4 & 1.1 & \textbf{1.0} \\
\hline
\textbf{Rotation Perturbations} & & & & & & & \\
5° & 0.49 (0.51\reddown) & 23.13& \textbf{0.86} (\textbf{0.01}\reddown) &\textbf{0.93} (\textbf{0.01}\reddown)&51.1 &1.1 &\textbf{1.0}\\
10° &  0.49 (0.51\reddown)& 22.72& \textbf{0.86} (\textbf{0.01}\reddown) &\textbf{0.93} (\textbf{0.01}\reddown)&51.4& 1.1 &\textbf{1.0}\\
15° & 0.48 (0.52\reddown)& 22.53& 0.86 (0.01\reddown) & \textbf{0.94} (\textbf{0.00})&51.9 &1.1 &\textbf{0.0}\\
\hline
\textbf{Elastic Deformation} & & & & & & & \\
Low  & 0.82 (0.18\reddown) & 31.05 & 0.85 (0.02\reddown)& \textbf{0.93} (\textbf{0.01}\reddown) &18.0&2.3 &\textbf{1.0}\\
Medium  & 0.67 (0.33\reddown) & 27.64& 0.85 (0.02\reddown)& \textbf{0.93} (\textbf{0.01}\reddown) &33.5&2.3 &\textbf{1.3}\\
High  & 0.59 (0.41\reddown) & 26.25& 0.84 (0.03\reddown)& \textbf{0.92} (\textbf{0.02}\reddown) &41.2&3.4 &\textbf{2.1}\\
\hline


\end{tabular}
\begin{tablenotes}
\item[*] The percentage drop for PSNR is not reported as the baseline value is infinite.
\end{tablenotes}
\end{threeparttable}
\end{table*}

\subsection{Ablation study}
~
To investigate the individual contributions of the restoration and denoising paths in Star-Diff, we conducted an ablation study by decoupling the two U-Nets and applying them independently during the staining translation process.
As shown in Table~\ref{tab:ablation_study}, using either path alone leads to inferior performance, highlighting the complementary roles of restoration and noise removal. These results underscore the necessity of jointly leveraging both pathways to achieve high-quality and diagnostically faithful IHC image generation.

\begin{table*}[t]
\centering
\caption{Ablation Study of sampling paths. } 
\label{tab:ablation_study}
\begin{tabular}{c c| c c| c c} 
\toprule
Restoration Path& Denoise Path & SSIM &PNSR(dB)&Accuracy&SFS\\ 
\hline
$\checkmark$&$\checkmark$&0.53&21.30&0.68&0.74\\ \hline
$\checkmark$&&0.30 (0.23\reddown)&17.46 (3.84\reddown)&0.62 (0.06\reddown)&0.68 (0.06\reddown)\\ \hline
&$\checkmark$&0.39 (0.14\reddown)&15.77 (5.53\reddown)&0.64 (0.04\reddown)&0.68 (0.06\reddown)\\ 
\bottomrule
\end{tabular}
\end{table*}

\section{Discussion}
\label{sec:discussion}
~ 

\subsection{The Clinical Applicability of our work}
 
Our proposed Semantic Fidelity Score (SFS) offers a clinically aligned evaluation strategy by quantifying the preservation of diagnostic information, rather than low-level pixel similarity. This makes it broadly applicable to medical image generation tasks beyond pathology, such as radiology synthesis and biology structure generation \cite{Guo2025Vessel, liu2025treatment}.

In clinical settings, the Star-Diff framework is not limited to the translation of HER2, but can be extended to other staining targets, such as from H\&E to CD10 \cite{liu2021unpaired}, or PAS \cite{de2021pas}. Furthermore, it could be adopted for time-sensitive workflows such as intraoperative frozen section analysis, where rapid and reliable pathological feedback is critical for surgical decision-making. Traditional staining protocols, especially IHC, are too time-consuming for such scenarios and are therefore rarely used intraoperatively. 
Combined with the robust SFS metric, it allows clinicians to instantly assess diagnostic relevance, potentially reducing turnaround time and improving patient outcomes.

\subsection{Limitations}
Despite the contributions, our work has some limitations.
Most prominently, while the proposed SFS metric offers clinically meaningful evaluation, it relies on a pretrained classifier requiring patch-level annotations. To support the research community, we release the pretrained classifier weights, allowing others to assess diagnostic relevance without the need for retraining. In future work, we plan to replace this step with a foundation model to reduce annotation requirements and enhance generalizability.


Further, as only one public dataset of paired H\&E–IHC patches is available at the time of writing, broader validation is limited. We are currently developing an internal paired H\&E–IHC dataset to further validate Star-Diff and benchmark it against other SOTA models.

\section{Conclusion}
\label{sec:conclusion}
~
In this work, we address the challenge of virtual staining from H\&E to IHC, where generating diagnostically meaningful IHC images remains non-trivial due to the need to preserve structural fidelity while modeling biological variability. Moreover, evaluating vIHC is complicated by inevitable spatial misalignment between H\&E and IHC slides, rendering traditional pixel-based metrics inadequate.
To tackle these challenges, we propose an integrated framework combining Star-Diff, a structure-aware diffusion model that reformulates staining translation as an image restoration task, and the Semantic Fidelity Score (SFS), a task-driven metric designed to assess diagnostic consistency. Star-Diff leverages dual pathways to balance structural preservation and biomarker diversity, while SFS provides robust evaluation under misalignment and classifier uncertainty.
Comprehensive experiments on the BCI challenge's dataset demonstrate that our approach outperforms SOTA performance across both visual fidelity and diagnostic relevance, offering a practical and clinically meaningful solution for virtual IHC synthesis. Star-Diff ranks first on the challenge's leaderboard.
In clinical contexts, Star-Diff provides a reliable and rapid virtual staining solution by generating IHC images within seconds. This significantly reduces processing time while preserving essential molecular biomarker information. Such capability is particularly valuable in intraoperative workflows, where timely and accurate decision-making is critical. By enabling fast and diagnostically consistent virtual IHC synthesis, Star-Diff holds promise for improving turnaround time and enhancing patient outcomes during surgery.

\section{Acknowledgment}
\label{sec:acknowledgment}
This work was supported by the BMBF-funded SATURN3 project (01KD2206B; 01KD2206E) and the IMI BIGPICTURE project (IMI945358). The authors thank Reza Nasirigerdeh for his valuable proofreading support.

\printcredits

\bibliographystyle{cas-model2-names}
\bibliography{cas-refs}





\end{document}